# Evolution and Reflection of Ray-like Excitations in Hyperbolic Dispersion Media

*Hanan Herzig Sheinfux, Matteo Ceccanti, Iacopo Torre, Lorenzo Orsini,*

*Frank H.L. Koppens*[*]

E-mail: frank.koppens@icfo.eu

ICFO-Institut de Ciencies Fotoniques, 08860 Castelldefels (Barcelona), Spain

*Frank H.L. Koppens*[*]
ICREA-Institució Catalana de Recerca i Estudis Avançats, 08010 Barcelona, Spain

*Minwoo Jung[2], Gennady Shvets[3]*
Department of Physics, Cornell University, Ithaca, New York, 14853, USA
School of Applied and Engineering Physics, Cornell University, Ithaca, New York 14853, USA

Abstract: Experiments and simulations show hyperbolic dispersion materials support nanoscale ray-like excitations, but only partial understanding currently exists of how these nanorays form and evolve. We provide an analytical framework to understand these nanorays and show their behavior combine properties of classical rays and of waveguide modes.

1. Introduction

Polaritons in hyperbolic dispersion media are the exception to several unwritten rules of nanophtoonics. Where typical polaritons exhibit strong wavelength contraction only at a narrow bandwidth, with the maximum momentum limited by absorption, hyperbolic dispersion media (HyM) can support 3D excitations exhibiting strong wavelength contraction over a large bandwidth and which attain an almost unlimited momentum even in the face of absorption. However, this large momenta is restricted to a limited angular range and light tends to acquire a ray-like character inside a HyM. Physically, hyperbolic dispersion manifests in highly anisotropic systems, when the in-plane permittivity has a different sign from the out-of-plane permittivity. This type of strong anisotropy has been demonstrated to occur in in anisotropic plasma[1] and man-made metamaterials[2,3]. More recently, it has been shown certain crystals can exhibit very different phononic or plasmonic resonances properties for different axes[2–6]. In particular for materials such as isotopically pure hexagonal boron nitride (hBN), hyperbolic excitations with exceptionally low absorption have been observed and the experimental accessibility of these low-loss hyperbolic excitations has opened the way for extensive research on the nature of hyperbolic light and its propagation.
Recent research of hBN has brought to front the ray-like character of excitations in HyM. Nano rays are important in studies of small hBN nanogranules[7] and cavities[8], but are most commonly discussed in the context of planar HyM slabs. For this geometry, the existence of nano-ray excitations can be demonstrated numerically and is directly observable in experiments[9,10]. However, the theoretical effort to understand these ray-light excitations has been surprisingly limited and there is no complete description of how such rays form, propagate and when or how they eventually dissipate. More specifically, while intuitively we might expect ray-like excitations to behave in fundamentally different ways then ordinary modes, no such behavior has been identified except for the sharply localized shape of the wavefront.

Here, we provide an analytical framework to study ray-like excitations in slabs of HyM material and utilize it to explore a uniquely ray-like reflection mechanism. Specifically, we demonstrate a dipole source near the HyM emits rays with a nearly-Lorentzian profile which propagate in the HyM in a zig-zagging motion, set with a frequency dependent angle. These rays broaden as they propagate (due to absorption rather than dispersion) and are shown to acquire a phase in a discrete fashion, by reflection events from the top and bottom interfaces of the HyM. We then study the reflection of such a beam from the interface at which the substrate underneath the HyM changes from dielectric to metallic. Ordinarily, some degree of reflection can be expected from impedance mismatch considerations. However, we find that if the ray is incident exactly at the corner of the metallic substrate, the reflection strength is enhanced due to the limited overlap between the ray-like excitation and the modes of a HyM on a metal substrate. More precisely, the amplitude transmitted across the interface is shown to scale inversely with the width of the incoming ray.

This text is divided in the following manner: section 2 introduces the notation and the modes of a HyM slab; section 3 studies the formation and evolution of a ray-like excitation; section 4 covers the reflection of this ray at a substrate-change interface.

To focus the discussion, we will consider flakes of hexagonal boron nitride (hBN) to be the HyM media. hBN is a natural layered crystal which supports very well studied hyperbolic phonon polaritons (PhPs) and, especially so in isotopically pure hBN, exhibits remarkably long PhP propagation lengths[11].

2. Phonon polaritons modes

We first calculate the normal electromagnetic modes in a HyM slab of thickness $t$ (infinitely extended in $x, y$) on top of a dielectric substrate, with extended derivation appearing in the SI. Starting from Maxwell's equations, we derive the Helmholtz equation for TM polarized waves in anisotropic media at the quasi static limit, (i.e. assuming the vacuum momentum is much smaller than typical momentum of the excitations considered),

$$\frac{1}{\epsilon_{zz}(\omega;z)}\partial_x^2 E_x(x,z) + \frac{1}{\epsilon_{xx}(\omega;z)}\partial_z^2 E_x(x,z) = 0. \quad [1]$$

The dielectric permittivity is assumed to have the following structure

$$\epsilon_{xx}(\omega;z) = \begin{cases} 1 & z > t \\ \epsilon_x & 0 < z < t \\ \epsilon_s & z < 0 \end{cases}$$

$$\epsilon_{zz}(\omega;z) = \begin{cases} 1 & z > t \\ \epsilon_z & 0 < z < t \\ \epsilon_s & z < 0 \end{cases} \quad [2]$$

With $\epsilon_x, \epsilon_z$ the HyM (hBN) in plane and out of plane permittivity and $\epsilon_s$ for the (isotropic) substrate. We solve by anzats and find modes $A_{n \geq 0}$ of the form

$$\vec{E}_n(\vec{r}) \cdot \hat{x} = N_n e^{iq_n x}\psi_n(z)$$

$$\psi_n(z) = \begin{cases} t_2 e^{-q_n(z-t)} & z > t \\ e^{ik_n z} + r e^{-ik_n z} & 0 < z < t \\ t_1 e^{q_n z} & z < 0 \end{cases}, \quad [3]$$

with $q_n$ being the x-component of the nth mode's momentum, $k_n$ the (complex) z-component of the wavevector and $N_n$ a normalization coefficient. $r$ and $t_{1,2}$ are determined from the boundary conditions,

$$r = \frac{i\theta\epsilon_x+1}{i\theta\epsilon_x-1}, \quad t_1 = \frac{2}{i\theta\epsilon_x+1}, \quad t_2 = e^{ik_nt} + re^{-ik_nt}. \tag{4}$$

$\theta \equiv \sqrt{-\frac{\epsilon_z}{\epsilon_x}} = \theta_r + i\theta_i$ is an important coefficient, as it also connects $q_n, k_n$ as $q_n = k_n\theta$ (in the quasi static limit). From the same boundary conditions, we also directly obtain the resonance condition for the PhP mode in the slab,

$$k_n = (\pi n + i\log(r))/t, \tag{5}$$

with $n \geq 0$ an integer. Accordingly, there is an infinite number of modes exists in a finite slab of thickness $t$. The modes are equally spaced, but are not simple harmonics. We can identify $q_n = q_0 + \Delta$, with $q_0 = \theta \log(r)/t$, $\Delta = \theta\pi/t$.

Finally, we redefine (in SI) a bi-orthogonal basis denoted by $\Xi_n$ so that $\int \psi_n \Xi_m dz = \delta_{mn}$. That is, bi-orthogonal with respect to $dz$ integration, rather then volume integration. Very similar modes occur for hBN over a metallic susbtrate, except that the modes have a node at $z = 0$, due to screening, and that the first mode of the slab is the $n = 1$ mode with

$$k'_{n\geq 1} = (\pi(2n-1) + i\log(r))/2t. \tag{6}$$

3. Modal ray formation and propagation

While the above modes are the basic modes of the system, numerical and experimental studies show that placing a dipole (or, simialrly, a near field microscope tip) near to an hBN slab launches excitations resembling nanorays. Of course, this ray-like excitation is really a superposition of the basic modes we found for the uniform slab. Indeed, if we consider the power coupled from the dipole to each of the modes, we obtain (with some minor approximtions) a geometric series, with the ratio between terms being $e^{i\Delta(z\pm\theta x)}$. Summing the series gives

$$\psi(x, 0 < z < t) = C \frac{e^{ik_0(z-t+\theta x)}}{1 - Le^{i\Delta(z-t+\theta_r x - i\theta_i h)}} + \frac{e^{-ik_0(z-\theta x)}}{1 - Le^{-i(z-t-\theta_r x - i\theta_i h)}} \tag{7}$$

with

$$L = \exp(\Delta\theta_i x - \Delta\theta_r h), \quad C = \frac{t_1}{\sqrt{2t}} e^{q_0(it+\theta h)} A \tag{8}$$

and $A$ a normalization factor.

Notably, the loss factor $L$ is a function of the propagation distance not of $z$, and for typical parameters is close to unity magnitude (i.e. for $x, h$ that are not too large and for most frequencies inside the restrahlen band of isoptopic hBN). We can identify the two terms on the right of equation [24] as the upward and downwards propagating components of the ray (respectively). The upwards component of the ray obtains it's maximum when $z_+ = z - \theta_r x$ is equal to $pt$, for $p$ an even integer and the downwards propagating component when $z_+ = z - \theta_r x$ is equal to $pt$, for $p$ an odd integer. That is, for every $x$, there is one integer $p$ so that one of the component peaks inside the flake (for $0 < z < t$). Expanding the upwards propagating (downwards propagating) component around $z_+$ ($z_-$), we get

$$\psi(x,z) = \sum_p C\left[r^p e^{-iq_0 z_+} \Delta \frac{\frac{\Gamma}{2} + i\Delta^{-1}(z_+ - pt)}{\left(\frac{\Gamma}{2}\right)^2 + (z_+ - pt)^2} + r^{p+1} C e^{iq_0 z_-} \Delta \frac{\frac{\Gamma}{2} - i\Delta^{-1}(z_- - pt)}{\left(\frac{\Gamma}{2}\right)^2 + (z_- - pt)^2}\right], \tag{9}$$

where $\Gamma = 2(1-L)\Delta^{-1} \simeq 2\theta_i x + 2\theta_r h$ is the width of the Lorentzian. $2\theta_i x$ can be thought of as the total amount of absorption the beam undergoes (where for realistic material parameters $\theta_i \ll 1$). Hence, in contrast with gaussian beams in ordinary media, the nanorays in hyperbolic media broaden thanks to absorption rather then dispersion. This is because the planewave components constituing the nanoray are co-propagating with almost the same angle, but not the same rate of absorption. Higher order modes are absorbed faster and decay, reducing the momentum of the beam and broadening it.

The ray-like components of eq. [9] move along a diagonal defined by $z_\pm = z - \theta_r x = pt$. Tracking them, we see that when an upwards propagating compoent reaches the top of the flake, a downwards propagating one (the $p+1$ component) emerges as though reflected (and vice versa when the downwards propagating component reaches the bottom). This creates the zig-zagging ray shown below, as expected for a ray.

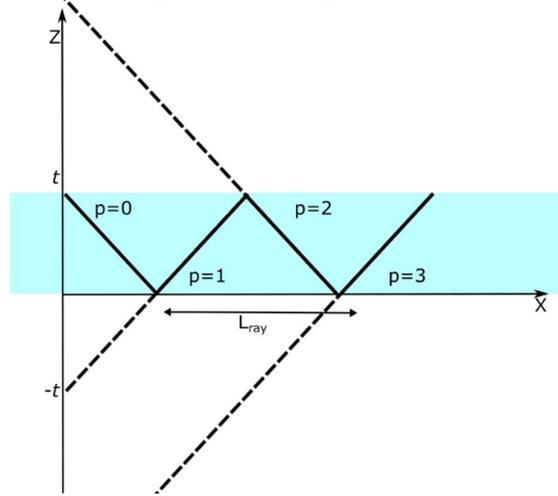

**Figure 1,** illustration of ray zig zagged propagation

Notably, while the nanoray (eq. [8]) has a spatially dependent phase, $e^{\mp i q_0 z_\pm}$, the center of the beam (where $z_\pm = pt$) has a practically stationary phase, as befitting a ray-like excitation. However, whenever the nanoray is reflected from the top or bottom interface, the reflected component is multiplied by the reflection coefficient, $r$. This lends to a discrete phase jump of $imag[\log(r)]$ at every reflection event. Furthermore, since $|r| < 1$, the amplitude of the ray also drops down at every reflection event, representing a discrete absorption mechanism. The accumulative effect of these discrete phase jumps, corresponds to a total rate of phase accumulation of $k_0 = 2\log(r)/L_{ray}$, resembling the first PhP mode.

Building on these insights, we can further simplify the ray to the form inside the flake (except for the vicinity of the flake top and bottom interface, where two such components contribute),

$$\psi\left(\frac{pt}{\theta_r} < x < \frac{t(p+1)}{\theta_r}, z\right) = r^p e^{\pm i k_0 (z \pm (\theta_r x - pt)) - \theta_i k_0 x} f(z_\pm; \Gamma), \quad [10]$$

$f(z_\pm - 2mt, \Gamma)$ is a Lorentzian profile function of width $\Gamma(x)$ that propagates at a $\pm \text{atan}(\theta_r)$ angle to the x-y plane,

$$f(0 < z < t, x; \Gamma) = N e^{q_0 \theta h} \frac{\frac{\Gamma}{2} + i\Delta^{-1} z}{\left(\frac{\Gamma}{2}\right)^2 + z^2}. \quad [11]$$

The SI shows the complete details of the calculations, and also demonstrates the ray decays non-exponentially outside of the slab. Furthermore, it is also shown that the exact profile of the ray, $f(z, x; \Gamma)$, depends on the details of the excitation profile. Intuitively, the ray profile tends to more strongly localized if more energy couples to higher order (higher momentum) modes.

We note that optical absorption plays three roles in the propagation of the beam. First, by broadening the beam, as explained above. Second, throught the $e^{-q_0 \theta_i x}$ term, which causes a slow uniform decay of the overall ray amplitude. Third, through the reflection coefficient – whenever the nanoray bumps on one of the interfaces, since $|r| < 1$, it loses some amplitude. Thus, the beam amplitude is decreasing in discrete jumps, similarly to its phase. As shown in Fig. 3d, these reflection losses are frequency dependent and very sensitive to the permittivity of environment (i.e. substrate\superstrate identity). Remarkably, reflection losses amount for a significant portion of the total absorption, comparable and possibly exceeding the direct losses incurred through propagation. However, controlling the substrate outside of the hBN can reduce these losses considerably. We suggest this effect is in the root of the increased propagation lengths observed in suspended hBN flakes, compared to hBN on an SiO2 substrate, as even a small change in permittivity can induce a substantial increase in reflection losses. Alternatively, one can also replace the substrate with a metallic one. The results on ray formation on a metallic substrate are included in the SI and show that a similar ray exists in this case, but that due to metallic screening it is reflected with a $\pi$ phase shift at the substrate. This is again in agreement with the recent experimental of extremely high quality factors in hBN on a metalized substrate[12].

4. multimodal ray reflection and transmission

Having laid down the framework to describe nanorays in hyperbolic media in the above section, we can now use this framework and explore the nanoray's behavior. In particular, the anomalous reflection of the ray due to abrupt substrate changes.

An essential step to obtain the reflection is calculating the the overlap in electric (and magentic) field between the ray coming to the substrate-change interface (from the dielectric side) and the modes it can transmit to on the metallic-substrated side. Intuitively, we can expect the field overlap to be minimal when the ray is incident at $z = 0$, when the PhP modes on the metallic substrate sides have a node. This impression can be directly substantiated by numerically integrating the field profile, as demonstrated in the figure below (for a realistic isotopically pure h$^{10}$BN flake).

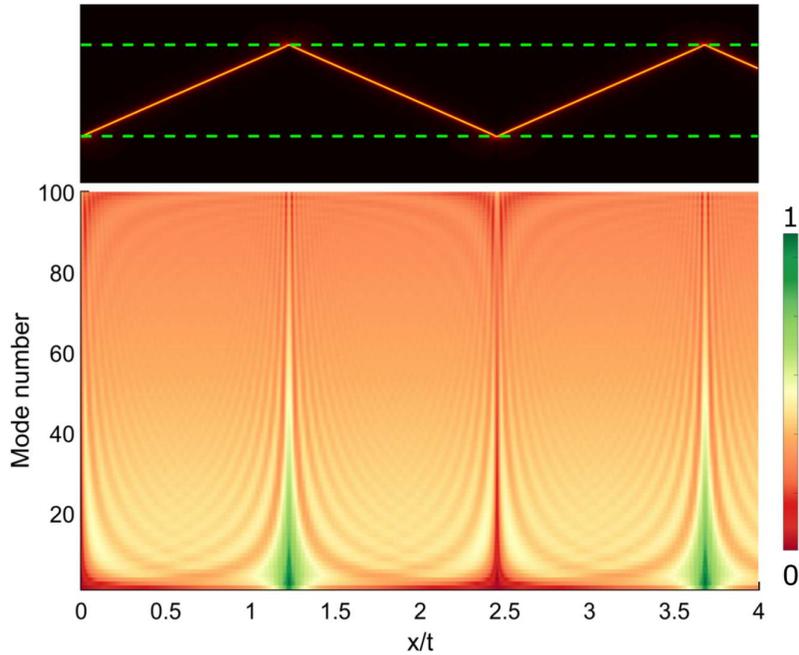

**Figure 2,** calculated electrical field overlap between the ray and the PhP modes on a metallic substrate. The map is of the overlap integral as a function of the mode number on the metal side (starting from $n = 1$) and the distance the multimodal ray transversed in $x$ (i.e. assuming this is the $x$ at which the substrate changes from dielectric to meallic). Notably, the overlap integral has a strong minimum when the ray is centered on the metallic corner at $z = 0$, except for large mode numbers ($M > 60$) when the width of the beam is comparable to the typical scale of the mode's profile. Broadening by loss in this case is seen to be very small.

The dependence of this overlap integral on the beam width Γ strongly depends on the beam profile. The beam profile in eq [9] has a complicated dependence on Γ, but for other ray profiles (i.e. for other excitation profiles) a power-law dependence can be extracted and, when $\Gamma \ll t$ and the beam is incident on the corner precisely, we find

$$\int_{-\infty}^{\infty} \psi_i \xi'_n dz = \eta\left(\frac{\Gamma}{t}\right), \qquad [12]$$

where $\eta \sim \left(\frac{\Gamma}{t}\right)^{3.5}$ rapidly vanishing when $\Gamma \to 0$.

The continuity of $E_z, B_y$ at the interface, implies that $\partial_z E_z, \partial_z B_y$ are also continuous, so that $\partial_x \psi, \psi$ are continous on both sides of the virtual interface defined by the change in the substrate. We signify with $\psi_i, \psi_r$ and $\psi_t$ the incident reflected and transmitted components. In order for the boundary condition on $\partial_x \psi_i + \partial_x \psi_r$ to cancel we can expect the phase of the reflection coefficient so that $\partial_x \psi_i \approx +\partial_x \psi_r$. More explicitly, we define the beam reflected from the interface as

$$\psi_r\left(x - \frac{w}{2}, z\right) = \psi_i\left(\frac{w}{2} - x, z\right) - \sum_n \delta_n \psi_n, \qquad [13]$$

where $\sum_n \delta_n \psi_n$ a correction term to the reflection beam (also propagating in the $-x$ direction). Next, we will show the magnitude of $\sum_n \delta_n \psi_n$ is small, and for some ray profiles decays with $\frac{\Gamma}{t}$. In this notation, the boundary conditions are

$$\begin{aligned} 2\psi_i - \sum_n \delta_n \psi_n &= \sum_m t_n \psi'_n \\ \sum_n \delta_n q_n \psi_n &= \sum_n t_n q'_n \psi'_n \end{aligned} \qquad [14]$$

from which we obtain a full set of linear equations to resolve $\delta_n$, in which all of the constant terms are proportional to $\eta$. Therefore, for critical incidence, when $\eta$ has a minimum, so does the reflection (and accordingly the transmission). As mentioned above, for certain ray profiles $\eta$ scales with a power of $\Gamma$ and the magnitude of all the $\delta_n$ coefficients vanishes when $\Gamma \to 0$. As is evident from the second boundary condition, the amplitude of the transmitted modes, $t_n$, is comparable in magnitude to the $\delta_n$ coefficients. Accordingly, the power transmitted through the virtual interface has a minimum when the beam is incident exactly at the metallic corner and can even disappear in the $\Gamma \to 0$ limit. We emphasize the enhanced reflection (reduced transmission) when the beam is incident at the metallic corner is an interference effect, where the PhP modes constituting the ray couple to modes on the metallic substrate in a destructively interfering fashion. At the same time, we note an opposite effect, enhanced transmission, is expected to occur when the beam is incident far away from the metallic corner (a few times $\Gamma$ or more).

In conclusion, we established an analytical framework to describe ray-like excitations in hyperbolic media and started upon studying the unique behavior of these excitations. Most intriguingly, these rays combine some aspects of their constituent modes (e.g. phase accumulation) with a ray-like behavior (e.g. discretized phase accumulation) and especially so in the transmission and reflection of the ray at substrate-change interfaces.

# Supplementary information

## Section 1 Phonon polariton modes

Taking the curl of Faraday's law of induction acting on an electric field at a fixed frequency $\omega$,

$$\vec{\nabla} \times (\vec{\nabla} \times \vec{E}) = \vec{\nabla}(\vec{\nabla} \cdot \vec{E}) - \nabla^2 \vec{E} = \frac{\omega^2}{c^2} \bar{\bar{\epsilon}} \cdot \vec{E}. \quad [15]$$

Here, $\vec{E}(\vec{r})$ is the vector amplitude of the $\omega$ frequency component of the electric field and $\bar{\bar{\epsilon}}$ is the permittivity matrix.
Substituting Ampère–Maxwell law, we obtain

$$\nabla^2 \vec{E} + \frac{\omega^2}{c^2} \bar{\bar{\epsilon}} \cdot \vec{E} = \vec{\nabla}(\vec{\nabla} \cdot \vec{E}). \quad [16]$$

We limit the discussion to waves in transverse magnetic (TM) polarization which propagate in the $x$-direction (so that the electric field probes the out of plane anisotropy). Hence, $B_x = B_z = 0$, $\partial_y = 0$. The $x$ component reads

$$\partial_z^2 E_x - \partial_x \partial_z E_z + \frac{\omega^2}{c^2} \epsilon_{xx} E_x = 0. \quad [17]$$

and

$$\frac{\epsilon_{xx}}{\epsilon_{zz}} \partial_x^2 E_x + \partial_z^2 E_x + k_0^2 \epsilon_{xx} E_x = 0. \quad [18]$$

Where $k_0 = \omega/c$.
In the quasi-static limit, assuming $k_0$ is much smaller than the typical momentum of $E_x(x,z)$, we obtain the anisotropic-medium Helmholtz equation:

$$\frac{1}{\epsilon_{zz}(\omega;z)} \partial_x^2 E_x(x,z) + \frac{1}{\epsilon_{xx}(\omega;z)} \partial_z^2 E_x(x,z) = 0. \quad [19]$$

We consider solutions of the form

$$\vec{E}_n(\vec{r}) \cdot \hat{x} = N_n e^{iq_n x} \psi_n(z)$$
$$\psi_n(z) = \begin{cases} t_2 e^{-q_n(z-t)} & z > t \\ e^{ik_n z} + r e^{-ik_n z} & 0 < z < t \\ t_1 e^{q_n z} & z < 0 \end{cases}, \quad [20]$$

with $q_n$ being the x-component of the nth mode's momentum, $k_n$ the (complex) z-component of the wavevector and $N_n$ a normalization coefficient. $r$ and $t_{1,2}$ are, as of yet undetermined, complex variables.
To simplify the notation, we use $\epsilon_x, \epsilon_z$ to signify hBN's anisotropic permittivity and $\epsilon_s$ for the (isotropic) substrate,

$$\epsilon_{xx}(\omega;z) = \begin{cases} 1 & z > t \\ \epsilon_x & 0 < z < t \\ \epsilon_s & z < 0 \end{cases}$$
$$\epsilon_{zz}(\omega;z) = \begin{cases} 1 & z > t \\ \epsilon_z & 0 < z < t \\ \epsilon_s & z < 0 \end{cases}. \quad [21]$$

Neglecting retardation effects, we have

$$q_n = \theta k_n, \quad [22]$$

with $\theta = \sqrt{-\frac{\epsilon_z}{\epsilon_x}}$.

Using the boundary conditions for the tangential electric and magnetic field and the relation $\partial_z B_y = \epsilon_x \omega E_x$ (Maxwell's equations), we get $\psi_n$ and $\epsilon_{xx} \omega \int \psi_n dz$ should be continuous at the interface, so that

$$t_2 e^{ik_n t} = e^{ik_n t} + re^{-ik_n t}$$
$$\frac{1}{i\alpha_n} e^{ik_n t} = \frac{\epsilon_z}{k_n}\left(e^{ik_n t} - re^{-ik_n t}\right)$$
$$1 + r = t_1 \quad [23]$$
$$\frac{\epsilon_z}{k_n}(1 - r) = \frac{1}{i\alpha_n}$$

From these equations we obtain Fresnel's reflection and transmission coefficients,

$$r = \frac{iq_n \epsilon_x - k_n}{q_n \epsilon_x + k_n} = \frac{i\theta\epsilon_x - 1}{i\theta\epsilon_x + 1}$$
$$t_1 = \frac{2k_n}{iq_n\epsilon_x + k_n} = \frac{2}{i\theta\epsilon_x + 1}, \quad [24]$$

and
$$t_2 = e^{ik_n t} + re^{-ik_n t}, \quad [25]$$

as well as an additional relation, known as the resonance condition
$$1 = r^2 \cdot \exp(2ik_n t). \quad [26]$$

Using the notation $\rho = \rho_r + i\rho_i = i \cdot \log(r)$, this gives
$$Re\{k_n\} = (\pi n + \rho_r)/t$$
$$Im\{k_n\} = \rho_i/t \quad [27]$$

So we can define,
$$k_n = k_0 + \Delta n. \quad [28]$$

Most notably, these $k_n$'s (and correspondingly the $q_n$'s also) are not proportional to $n$, since generally $\rho_r \neq 0$. In fact, the $n = 0$ mode tends to have a wavelength a few times larger than the wavelength of the $n = 1$ mode (at some frequencies, even an order of magnitude larger). Though higher order modes approach harmonics of each other rather fast (see also Fig. 1 of the main paper).

$$\xi_n = N_n e^{iq_n x}\begin{cases} t_1^c e^{-q_n(z-t)} & z > t \\ e^{-i\phi q_n z} + r^c e^{i\phi q_n z} & 0 < z < t, \\ t_1^c e^{q_n z} & z < 0 \end{cases} \quad [29]$$

where
$$r^c = \frac{i\epsilon_x \theta + 1}{i\epsilon_x \theta - 1} = r^{-1}$$
$$t_1^c = \frac{21}{i\epsilon_x \theta - 1} = t_1^\star \quad [30]$$

$\xi_n$ can be thought of as $\psi_n^\star(q_n \to q_n^\star)$, wher the waves still decaying evanescently away from the hBN, giving the different transmission and reflection coefficients.

The bi-orthogonal product is then defined as
$$\int \psi_n \xi_m d\bar{x} = \delta_{mn}. \quad [31]$$

Accordingly, we can now obtain the normalization coefficient by integrating,

$$N_n^{-2} = \int_{-\infty}^{\infty} \psi_n \xi_m \, dz = -\frac{|t_1|^2 + |t_2|^2}{2\alpha_n} + \int_0^t 1 + re^{-ik_n 2z} + r_c e^{ik_n^\star 2z} + rr_c \quad [32]$$
$$= 2t + \frac{r(e^{ik_n 2t}-1)}{2\phi q_n} - \frac{r^{-1}(e^{-ik_n 2z}-1)}{2\phi q_n} - \phi\frac{|t_1|^2+|t_2|^2}{2iq_n} \quad [33]$$
$$= 2t + P_n \quad [34]$$

In addition to this bi-orthogonal product, it is of prime interest to consider the product of integration on the $z - y$ plane, for example at the interface of the interior and exterior of the cavity.

$$\int \psi_n \xi_{m \neq n} dz = \int_t^\infty (e^{i\phi q_n t} + re^{-i\phi q_n t})(e^{-i\phi q_m t} + r^{-1} e^{i\phi q_m t}) e^{-(q_n - q_m)(z-t)} dz + \int_{-\infty}^0 t_1 t_1^c e^{(q_n + q_m)z} dz +$$
$$+ \int e^{i\phi q_n z - i\phi q_m z} + re^{-i\phi q_n z - i\phi q_m z} + r^{-1} e^{i\phi\ nz + i\phi\ mz} + e^{-i\phi q_n z + i\phi q_m z} dz \quad [35]$$
$$= \frac{1}{q_n + q_m}\left(t_1 t_1^c - e^{i\phi(q_n - q_m)t} - e^{-i\phi(q_n + q_m)t + i\rho} - e^{i\phi(q_n + q_m)t - i\rho} - e^{-i\phi(q_n - q_m)t}\right) +$$

$$\frac{1}{q_n+q_m}\frac{1}{i\phi}\left(e^{i\phi(q_n-q_m)t} - e^{-i\phi(q_n-q_m)t}\right) + \frac{1}{q_n-q_m}\frac{1}{i\phi}\left(re^{-i\ (q_n-q_m)t} + r^{-1}e^{i\phi(q_n-q_m)t}\right)$$
[36]

$$= \frac{1}{q_n+q_m}\left(|t_1|^2 - 2\cos((k_n-k_m)t) - 2\cos((k_n+k_m)t-\rho)\right) +$$
$$\frac{1}{q_n+q_m}\frac{2}{i\phi}\sin((k_n-k_m)t) + \frac{1}{q_n-q_m}\frac{2}{i\phi}\cos((k_n+k_m)t-\rho) \quad [37]$$

Substituting $k_n$,
$$= \frac{1}{q_n+q_m}\left(|t_1|^2 + (-1)^{m-n} - \cos((n+m)\pi - 3\rho)\right) + \frac{1}{q_n-q_m}\frac{1}{i\phi}\cos((n+m)\pi - 3\rho)$$
[38]

Clearly, this integral does not cancel out in the general case, though it is very near to cancelling. For example, numerically (credit matteo) we get the following values, for the $\int \psi_n \xi_m dz$ integral:

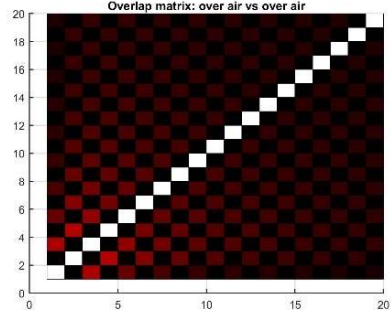

**Figure S1,** matrix of $\int \psi_n \xi_m dz$ as function of the mode numbers, $n, m$. White is maximum.

Using the Gramm-Schmidt process we reconstruct our modes so that the new base, denoted by $\Xi_n$, is bi-orthogonal relative to the $\int \psi_n \xi_m dz$ product:

$$\Xi_0 = \frac{\xi_0}{\int \psi_0 \xi_0 dz} \quad [39]$$

$$\Xi_1 = \frac{\xi_1}{\int \psi_1 \xi_1 dz} - \Xi_1 \int \psi_1 \xi_0 dz \quad [40]$$

...

$$\Xi_n = \frac{\xi_n}{\int \psi_n \xi_n dz} - \sum_{m<n} \Xi_m \int \psi_n m dz, \quad [41]$$

so that,
$$\int \psi_n \Xi_m dz = \delta_{mn}. \quad [42]$$

We denote the basis transform matrix between the $\xi_n$ and $\Xi_n$ bases is defined as
$$[\xi_n] = \bar{\bar{X}}[\Xi_m] \quad [43]$$

The general form of an excitation obtained in terms of these $\psi_n$'s is therefore
$$\psi_i = \sum_{n=0}^{\infty} u(x,n)\psi'_n \quad [44]$$

With $u(x,n)$ evolving as $e^{-iq_n x}$ in a pristine hBN.

## Section 2   Ray-like excitations

We consider nearfield excitations of the general form,
$$\psi = \sum_n u_n \psi_n. \quad [45]$$

In addition to far field modes (irrelevant to the current discussion) a dipole source above the HyM slab will project energy to all possible nearfield possible modes. Initially we consider a a dipole at $x = 0$, $z = t + h$, which induces an excitation with $u_n = t_2 e^{-\alpha_n h \pm ik_n t} = (-1)^n t_1 e^{-\alpha_n h \pm ik_n t}$. Allowing each mode to evolve in $x$ separately, with $e^{iq_n x} = e^{i\theta k_n x}$

$$\psi(x, 0 < z < t) = A\sum_n N_n e^{-k_n \theta h}(-1)^n t_1\left(e^{ik_n(z+\theta x)} + re^{-ik_n(z-\theta x)}\right) \quad [46]$$

with $A$ a normalization constant.

Using $k_n = k_0 + \Delta n$ with $\Delta t = \pi$ and $e^{ik_0 t} = r$, we get

$$\psi(x, 0 < z < t) = At_1 e^{ik_0(z+\theta x+i\theta h)} \sum_n N_n e^{in\Delta(z-t+\theta x+i\theta h)}$$
$$+ Art_1 e^{-ik_0(z-\theta x-i\theta h)} \sum_n N_n e^{-in\ (z-t-\theta x-i\theta h)} \quad [47]$$

This expression resembles a geometric series, except for the $n$-dependence in the normalization factor, which from [34] was shown to be to be $N_n = \sqrt{\frac{1}{2t+P_n}}$ with $P_n$ proportional to $\frac{1}{q_n}$ and accordingly $P_n \propto 1/n$. It follows that for large $n$, the normalization coefficient $N_n$ depends weakly on $n$. More specifically, $q_n = \frac{\pi n + \rho}{\theta t}$ and since $\theta$ is typically order one (or smaller), the evanescent decay length of $n \geq 1$ is smaller than flake thickness. Accordingly, we find that to a good approximation $N_{n \geq 1} = const$ under typical experimental conditions. This approximation is less accurate for $n = 0$, for which the mode leaks out more significantly outside of the flake, so that $P_0$ can be much larger then $P_{n>0}$. Based on these arguments, we will assume $N_0 = \frac{1-\nu}{\sqrt{2t}}$ and $N_{n>0} = \frac{1}{\sqrt{2t}}$, with $1 > \nu > 0$ a frequency dependent coefficient. We note the $N_0, N_{n>0}$ distinction is not very consequential for the purposes of this work. Furthermore, the size of $\eta$ will, in practice, depend on the experimental details (i.e. the coupling mechanism) as much as on the value of the normalization coefficient.

Under these assumptions, we obtain

$$\psi(0, 0 < z < t) \simeq \frac{1+r}{\sqrt{2t}} \left( A \frac{e^{k_0(\theta h + iz + i\theta x)}}{1 - e^{i\Delta(z-t-i\theta\ )}} + rA \frac{e^{-k_0(iz-\theta h - i\theta x)}}{1 - e^{-i\Delta(z-t-i\theta\ )}} - \nu \psi_0 \right), \quad [48]$$

or equivalently [48]

$$\psi(x, 0 < z < t) = \Delta C \left( r^{-1} \frac{e^{ik_0(z+\theta x)}}{1 - L e^{i\Delta(z-t+\theta_r x - i\theta_i h)}} + \frac{e^{-ik_0(z-\theta x)}}{1 - L e^{-i\Delta(z-t-\theta_r x - i\theta_i h)}} \right) - \tilde{\nu} \psi_0, \quad [49]$$

with

$$L = \exp(\Delta \theta_i x - \Delta \theta_r h), C = r \frac{1+r}{\Delta \sqrt{2t}} A e^{k_0 \theta h}, \tilde{\nu} = \frac{1+r}{\sqrt{2t}} \nu. \quad [50]$$

We focus on the case where $|1 - L| \ll 1$ and $z - \theta_r x$ is small. This requires that $h$ is sufficiently small, so $\theta_r \frac{\pi h}{t} \ll 1$ and that the propagation length is small compared to $\theta_i^{-1}$. Since in the bulk of the Restrahlen band, $\theta_i$ is generally small and this condition is realistic and experimentally viable. Notably, $L$ is independent of the substrate or superstrate, indicating it stems from the propagation of the ray and not tied with reflection losses.

The left term (right term) of eq [48] obtained it's maximal value, $\frac{1}{1-L}$, at $z_+ = z - t + \theta_r x$ (at $z_- = z - t - \theta_r x$) and also at $z_+ + mt$ for even $m$ (at $z_- + mt$ for odd $m$), for any $p \in \mathbb{Z}$. Expanding around these $z_\pm$, we get

$$\psi(x, z) \simeq \sum_{p \in \mathbb{Z}} \left[ \frac{r^{-1} \Delta C e^{ik_0(z+\theta x)}}{1 - L(1 + i\Delta(z_+ - pt))} + \frac{\Delta C e^{-ik_0(z+\theta x)}}{1 - L(1 - i\Delta(z_- - pt))} \right] - \tilde{\nu} \psi_0 \quad [51]$$

$$= \sum_p \left[ r^{-1} \Delta C e^{ik_0(z+\theta x)} \frac{1 - L - i\Delta(z_+ - pt)}{(1-L)^2 + \Delta^2 (z_+ - pt)^2} + \Delta C e^{-ik_0(z+\theta x)} \frac{1 - L + i\ (z_- - pt)}{(1-L)^2 + \Delta^2 (z_- - pt)^2} \right] - \tilde{\nu} \psi_0. \quad [52]$$

Which we can recast as

$$\psi(x, z) = \sum_p \left[ C e^{-ik_0(z_+ + t)} \frac{\frac{\Gamma}{2} - i(z_+ - pt)}{\left(\frac{\Gamma}{2}\right)^2 + (z_+ - pt)^2} + r^{-1} C e^{ik_0(z_+ + t)} \frac{\frac{\Gamma}{2} + i(z_- - pt)}{\left(\frac{\Gamma}{2}\right)^2 + (z_- - pt)^2} \right] - \tilde{\nu} \psi_0, \quad [53]$$

where $\Gamma = 2(1-L)\Delta^{-1} = 2\Delta^{-1} \left( \exp\left(\Delta \theta_i \frac{w}{2} - \Delta \theta_r h\right) - 1 \right) \simeq 2\theta_i x + 2\theta_r h$ is the FWHM of the Lorentzian distribution. Notably, $2\theta_i x$ can be thought of as the total amount of absorption the beam undergoes.

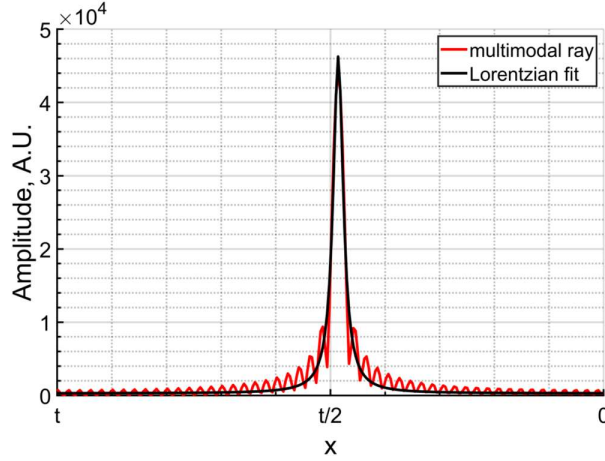

**Figure S2,** calculated profile of a multimodal ray. The red line shows the multimodal ray induced by a dipole above an hBN slab, with the calculation including the first 100 modes. The ray is allowed to propagate from $z = t$ to approximately the middle of the slab. The black lines shows good agreement with a Lorentzian fit.

Noting that that the maximum of the field is about $\Gamma$ times higher then the minimum, we omit when conveient the contribution of the $\eta$ term when describing the behavior of the ray. Clearly, the total excitation in this case can be thought of as the combination of a ray-like excitation (the object of this study) and a PhP mode of the lowest order.

To determine the value of the normalization constant, we return to equation [45] and consider the energy contained in the $u_n$ coefficients. At $x = 0$,

$$norm(\psi) = \sum |u_n|^2 = \sum |t_2|^2 e^{-2Re[q_n]h}$$
$$= |t_2|^2 e^{-Re[q_0 h]} \sum e^{n\theta \Delta h} = |t_2|^2 e^{-R\ [q_0 h]} \frac{2}{\Delta \Gamma} \qquad [54]$$

and it follows the normalization constant, $A$, is proportional to $\sqrt{\Gamma/t}$, so we can accordingly denote the coefficient multiplying the ray as

$$C = C_0 \sqrt{\frac{\Gamma}{t}}. \qquad [55]$$

For a final simplification of the ray form, we note that, since $e^{ik_0 pt} = r^p$, the $p^{th}$ component of the beam is similar to the $p - 2$ component up to an $r^2$ multiplicative factor. At the interfaces, $z = 0, t$, the field profile of the upward and downward propagating component is identical, except for an additional $r$ multiplicative factor (for the reflected component). We therefore obtain that inside the flake

$$\psi\left(\frac{pt}{\theta_r} < x < \frac{t(p+1)}{\theta_r}, z\right) = r^p e^{\pm ik_0(z\pm(\theta_r x-pt))-\theta_i k_0 x} f(z_\pm, \Gamma), \qquad [56]$$

with $f(z_\pm - 2mt, \Gamma)$ being the (Lorentzian) profile of the ray propagating in the upwards (+) or downwards (-) direction.

$$f(z, x; \Gamma) = \frac{A(r+r^2)}{\pi} \sqrt{\frac{t}{2}} e^{q_0 \theta h} \frac{\frac{\Gamma}{2}+i}{\left(\frac{\Gamma}{2}\right)^2 + z^2}. \qquad [57]$$

A similar expression is obtained when the ray is close to the edge of the flake, where

$$\psi\left(\frac{pt}{\theta_r} < x < \frac{t(p+1)}{\theta_r}, z\right) = e^{\pm ik_0(z\pm(\theta_r x-pt))-\theta_i k_0 x} \left(r^{p-1} f(z_\mp, \Gamma) + r^p f(z_\pm, \Gamma)\right), \quad [58]$$

or

$$\psi\left(\frac{pt}{\theta_r} < x < \frac{t(p+1)}{\theta_r}, z\right) = e^{\pm ik_0(z\pm(\theta_r x-pt))-\theta_i k_0 x} \left(r^p f(z_\pm, \Gamma) + r^{p+1} f(z_\mp, \Gamma)\right), \quad [59]$$

Depending on which interface the ray hits.

When the ray is close to the flake edges, The $e^{-\theta_i k_0 x}$ factor represents the continuous absorption of the ray in the hBN, which is physically distinct from the discrete absorption that the ray experiences at every reflection since $|r| < 1$. In similarity, the $e^{-\theta_i k_0 x}$ in eq. [60] also accounts for continuous phase accumulation at a $\frac{\theta_i \rho_i}{t}$ rate. However this phase accumulation mechanism is negligible, since $\theta_i$ and $\rho_i$ are both small for realistic parameters, so that $(\theta_i \rho_i)$ is typically two orders of magnitude longer then any other relevant length scale. Neglecting this mechanism, we arrive to the conclusion that phase accumulation in the nanoray is discrete, occurring via reflection events from at the top and bottom interfaces of the hBN.

For completeness, consider the decay of the ray outside of the hBN, for $z < 0$,

$$\psi\left(\frac{w}{2}, z \le 0\right) = \sum_n N_n t_1 e^{q_n(z+h)+iq_n x} = \sum_n N_n e^{\theta(k_0 + \Delta n)(z+h+ix)} \tag{61}$$

$$= e^{-q_0 \theta\left(h+z+i\frac{w}{2}\right)} \sum_n N_n e^{n\Delta\theta(z+h+ix)} \tag{62}$$

On the approximation that the normalization constant is $n$ independent, we get

$$\psi(x, z < 0) = \psi(x, z \to 0^+)\left(1 - e^{\Delta\theta(h+i\ )}\right) A \frac{e^{k_0 \theta(h+z+ix)}}{1 - e^{\Delta\theta(h+z+ix)}} \tag{63}$$

Here $\psi(x, z \to 0^+)\left(1 - e^{\Delta\theta(h+ix)}\right)$ is required by the continuity of the nanoray at $z = 0$. From this, we see the ray decays at a super-evanescent rate outside of the hBN slab. For $z \ll t$, it is proportional to $1/z$, whereas for $z \gtrsim t$, it is exponential, $e^{\theta k_0 z}$. Similar results can be obtained for the decay of the ray above the flake, for $z > t$.

For $\theta x = t$

$$\psi(x, z < 0) = \psi(x, z \to 0^+)\left(1 - e^{\Delta\theta(h+ix)}\right) A \frac{e^{k_0 \theta(h+z+i\ )}}{1 + e^{\Delta\theta(h+z)}} \tag{64}$$

## Section 3  Alternative ray profiles

The ray formalism above is naturally excited by a source that excites all modes with equal effiency. This represents a delta-localized dipole. A more realistic excitation, such as a nanoscattering objects (e.g. a sphere, a SNOM tip) is expected to launch more preferable to modes with momentum similar ot the spatial momentum of the scatterer. It is therefore interesting to note that different excitations profiles can also induce ray-like excitations, but with a different profile. A full analysis of the ray-profile as function of the excitation profile is beyond the scope of the current work, but to demonstrate the potential influence of the excitation on the ray profile, we consider a simple example where the excitation is of the form $u_n = n t_2 e^{-\alpha_n h \pm i k_n t}$. It can straightforwardly be shown that for this case

$$\tilde{\psi}(0, 0 < z < t) \simeq \frac{1+r}{\sqrt{2t}}\left(A \frac{e^{k_0(\theta h + i z + i\theta\ )}}{(1 - e^{i\Delta(z-t-i\theta\ )})^2} + rA \frac{e^{-k_0(iz - \theta h - i\theta\ )}}{(1 - e^{-i\Delta(z-t-i\theta h)})^2}\right). \tag{65}$$

And following a similar derivation, we get

$$\tilde{\psi}\left(\frac{pt}{\theta_r} < x < \frac{t(p+1)}{\theta_r}, z\right) = r^p e^{\pm i k_0 (z \pm (\theta_r x - pt)) - \theta_i k_0 x} \tilde{f}(z_\pm, \Gamma), \tag{66}$$

with

$$\tilde{f}(z, x; \Gamma) = \frac{A(r+r^2) e^{q_0 \theta h}}{\pi^2} \sqrt{\frac{t^3}{2}} \left(\frac{\frac{\Gamma}{2}+iz}{\left(\frac{\Gamma}{2}\right)^2 + z^2}\right)^2. \tag{67}$$

Clearly, this alternative excitation yields a much narrower (more localized) beam. The narrower profile is a consequence of the higher (average) momentum of the ray. To show this, we compare, below, the excitation profile $(u_n)$ for both types of excitations.

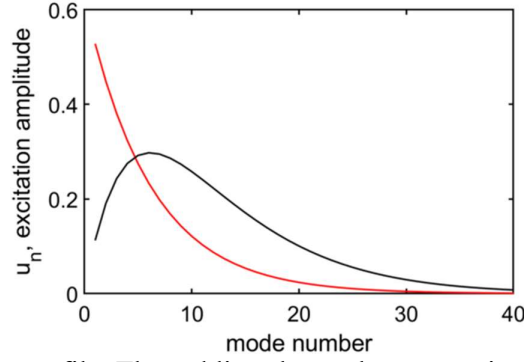

**Figure S3,** two excitation profile. The red line shows the $u_n$ required to create the Lorentzian distribution $f(z, x; \Gamma)$ described in previous sections. The black shows the $u_n$ required to create the narrower distribution $\tilde{f}(z, x; \Gamma)$. Calculation is for a 30nm thick $h^{11}BN$ flake at $1400 cm^{-1}$ and an excitation source located $5nm$ above the sample, using the normalization descirbed in the text.

## Section 4    Rays on a alternative substrate

Consider an hyperbolic hBN flake lying on a perfect metal substrate, with $\epsilon_s \to -\infty$ permittivity. The polaritonic modes are obtained similarly to the case of a dielectric substrate, except the field cannot penetrate into the metal,

$$\psi'_n = 2N'_n e^{iq'_n x} \begin{cases} t_2 e^{q'_n z} & z > t \\ e^{ik'_n z} - e^{-ik'_n z} & 0 < z < t \\ 0 & z < t \end{cases} \qquad [68]$$

Here $k'_n, q'_n$ play the the $z$ and $x$ components of the complex wavevector, as before, but due to the fixed $-1$ reflection from the metallic substrate, they are:

$$k'_n = \theta^{-1} q'_n = t^{-1}(\pi(2n-1) + \rho)/2 . \qquad [69]$$

With $t_1, \rho$ the same as for the dielectric substrate.

Note that for metal, since $\rho > 0$, it follows that $n \geq 1$ is the first allowed mode. Consequentially, the momentum of the fundamental PhP over a metallic substrate is significantly larger then over a dielectric substrate, i.e. that $q'_1 > q_0$.

Similarly to before, we find the biorthogonal basis for $\psi$ is

$$\xi_n = N'_n e^{iq_n x} \begin{cases} t_2^\star e^{-q'_n(z-t)} & z > t \\ e^{-ik'_n z} - e^{ik'_n z} & 0 < z < t \\ 0 & z < 0 \end{cases} \qquad [70]$$

where now

$$t_2 = e^{-ik'_n t} - e^{ik'_n t} = (-1)^n 2i\sqrt{r} . \qquad [71]$$

so that

$$N'^{-2}_n = \int_{-\infty}^{\infty} \psi'_n \xi'^\star_n dz = -\frac{|t_1|^2}{2\alpha_n} + \int_0^t 2 - e^{-ik'_n 2z} - e^{ik'^\star_n 2z}$$

$$= 2t - \frac{e^{ik_n 2t} + e^{ik'_n 2t} - 2}{2\phi q_n} - \frac{|t_1|^2 \phi}{2i q_n} = 2t + P'_n .$$

[72]

We obtain a multimodal ray from a dipole excitation in a similar manner as for the dielectric substrate:

$$\psi'(0, 0 < z < t) = A' \sum_{n>0} N'_n t_2 e^{-k'_n \theta h}$$

$$= -2iA'\sqrt{r}e^{k_1'(i(z+\theta x)-\theta h)}\sum_n N_n e^{\Delta(i(z-t+\theta)-\theta h)} +$$
$$2iA'\sqrt{r}e^{-k_1'(i(z-\theta x)+\theta h)}\sum_n N_n e^{-\Delta(i(z-\theta x)+\theta h)} \approx$$
$$iA'\sqrt{\frac{2r}{t}}\left(-e^{k_1'(i(z+\theta x)-\theta h)}\sum_n e^{\Delta(i(z-t+\theta x)-\theta h)} + e^{-k_1'(i(z-\theta x)+\theta h)}\sum_n e^{-\Delta(i(z-\theta x)+\theta h)} - \eta\psi_1'\right)$$

$$= \Delta C'\left(-\frac{e^{k_1'(i(z+\theta x)-\theta h)}}{1-Le^{\Delta(i(z-t+\theta_r x)-\theta_i h)}} + \frac{e^{-k_1'(i(z-\theta x)+\theta h)}}{1-Le^{-\Delta(i(z-\theta_r x)+\theta_i h)}}\right) - \eta\gamma'\psi_1' \quad [73]$$

where

$$L = \exp(\Delta\theta_i x - \Delta\theta_r h), C' = iA'\sqrt{\frac{2r}{t}}e^{k_1'\theta h}, \gamma = iA'\sqrt{\frac{2r}{t}}. \quad [74]$$

As before, we expand the exponent in the denominator and neglect the $\theta_i$ term, obtaining,

$$\psi'(0, 0 < z < t) = \Delta C'\left(-\frac{e^{ik_1'(z+\theta x)}}{1-L-\Delta i(z_+ - mt)} + \frac{e^{-ik_1'(z-\theta x)}}{1-L+\Delta i(z_- - mt)}\right) - \eta\gamma'\psi_1' \quad [75]$$

$$\psi'(0, 0 < z < t) = C'\left(e^{-ik_1'(z-\theta x)}\frac{\frac{\Gamma}{2}-i(z_- - mt)}{\left(\frac{\Gamma}{2}\right)^2 + (z_- - mt)^2} - e^{ik_1'(z+\theta x)}\frac{\frac{\Gamma}{2}+i(z_+ - mt)}{\left(\frac{\Gamma}{2}\right)^2 - (z_+ - mt)^2}\right) - \eta\gamma'\psi_1' \quad [76]$$

## Section 5  Reflection coefficient calculation

To continue with our calculation, we to need to obtain the overlap integral between $\partial_x\psi, \psi$ of the ray-like excitation and $\xi_n'$, the $n^{\text{th}}$ mode of hBN on metal. Intuitively, if the ray reaches the interface at $z = 0$, where $\xi_n'$ has a node, we can expect the overlap to be small. This can be verified by calculating the overlap numerically in the general case.

The exact dependence of the overlap integral on $\Gamma$ and $x$, depends on the details of the excitation, the values of $u_n$, the (frequency dependent) normalization coefficient and so forth. However, in some simplified cases, we can extract a simple power-law dependence.

To this end, we restrict the discussion the first $M_\Gamma$ modes, with $M_\Gamma$ being the largest $m \in \mathbb{N}$ for which $t \gg m\Gamma$, otherwise, the field of the modes (on the metallic substrate) varies on a scale comparable to beam profile. Under this assumption, the amplitude of the $\psi_n'$ modes immediately above the surface can be approximated as $\psi_n' \approx \kappa_n z$. As a further simplification, we ignore the $z > t$ part of the overlap integrals (where the Lorentzian has already decayed completely). The $z < 0$ part is trivial, due to the metallic boundary condition.

We denote by $c = 2\Gamma^{-1}(x - pt)$ the height (in $z$) of the ray from $z = 0$ and use $\zeta = 2\Gamma^{-1}z$. Since the ray is strongly localized (with $k_0 \ll 1/\Gamma$), we neglect the exponential dependence $e^{\pm ik_0 z_\pm}$. To simplify, we also neglect the slower decaying $\pm i(z_+ - pt)$ term in the nominators and the $\eta\psi_0$ component of the ray-like excitation. We find,

$$\int_{-\infty}^{\infty}\psi_i\xi_n'dz \approx \int_0^t\left(\frac{\frac{\Gamma}{2}c}{\left(\frac{\Gamma}{2}\right)^2 + (z_+ - pt)^2} + r^{-1}C\frac{\frac{\Gamma}{2}c}{\left(\frac{\Gamma}{2}\right)^2 + (z_- - pt)^2}\right)N_0'k_n zdz \quad [77]$$

Using the notation $d = t - \theta_r x$ to signify the height the beam is incident at,

$$= N_0'k_n C\int_0^t 2\Gamma^{-1}\left(\frac{z+d}{1+4\left(\frac{z+d}{\Gamma}\right)^2} - \frac{d}{1+4\left(\frac{z+d}{\Gamma}\right)^2} + r\frac{z-d}{1+4\left(\frac{z-d}{\Gamma}\right)^2} + r\frac{d}{1+4\left(\frac{z-d}{\Gamma}\right)^2}\right)dz \quad [78]$$

$$= N_0'k_n C_0\sqrt{\frac{\Gamma}{t}}\left(\frac{\Gamma}{2}\ln\left(\frac{1+\left(\frac{2}{\Gamma}(t+d)\right)^2}{1+\left(\frac{2}{\Gamma}d\right)^2}\right) + r\frac{\Gamma}{2}\ln\left(\frac{1+\left(\frac{2}{\Gamma}(t-d)\right)^2}{1+\left(\frac{2}{\Gamma}d\right)^2}\right) - d\left(\mathrm{atan}(2\Gamma^{-1}(t+d))+\right)+\ldots\right.$$

$$\left.\ldots + r\,\mathrm{atan}(2\Gamma^{-1}(t-d)) - (1+r)\,\mathrm{atan}(2\Gamma^{-1}d)\right). \quad [79]$$

$$= (\pi(2n+1) - \rho)(\eta + \gamma d) \quad [80]$$

With

$$\eta = N_0' k_n C_0 \left(\frac{\Gamma}{t}\right)^{1.5} \left(\ln\left(\frac{1+\left(\frac{2}{\Gamma}(t+d)\right)^2}{1+\left(\frac{2}{\Gamma}d\right)^2}\right) + r \ln\left(\frac{1+\left(\frac{2}{\Gamma}(t-d)\right)^2}{1+\left(\frac{2}{\Gamma}d\right)^2}\right)\right), \quad [81]$$

$$\gamma = N_0' k_n C_0 \sqrt{\frac{\Gamma}{t}} \left(\operatorname{atan}(2\Gamma^{-1}(t+d)) + r \operatorname{atan}(2\Gamma^{-1}(t-d)) - (1+r)\operatorname{atan}(2\Gamma^{-1}d)\right) \quad [82]$$

Note $N_0' k_n C \sim t$ and hence, $\eta$ scales at least as with $\left(\frac{\Gamma}{t}\right)^{1.5}$. Specifically, for $d = 0$ and $\Gamma \ll t$ we find $\eta$ scales as

$$\eta \sim \left(\frac{\Gamma}{t}\right)^{1.5} \ln\left(1 + \left(\frac{\Gamma}{t}\right)^2\right) \sim \left(\frac{\Gamma}{t}\right)^{3.5}. \quad [83]$$

Hence, for $\delta \to 0$ and low losses, sufficiently small ray width $\Gamma$ the overlap vanishes completely. We emphasize the dependence of the overlap integral on $\Gamma/t$ depends on the details of the excitation and can be faster or slower than this, depending on the exact excitation and so forth.

In a similar fashion, we can calculate the magnetic field overlap which requires the $\int_{-\infty}^{\infty} \partial_x \psi_i \partial_x \xi_m' dz$ integral. This integral is shown in the appendix but has a minimal value that does not depend on $\frac{\Gamma}{t}$ and as such is less important to the discussion ahead.

We signify with $\psi_i$, $\psi_r$ and $\psi_t$ the incident reflected and transmitted components. Since the total reflected field In order for the boundary condition on $\partial_x \psi_i + \partial_x \psi_r$ to cancel we can expect the phase of the reflection coefficient so that $\partial_x \psi_i \approx +\partial_x \psi_r$. More specifically, we define the beam reflected from the interface as

$$\psi_r\left(x - \frac{w}{2}, z\right) = \psi_i\left(\frac{w}{2} - x, z\right) - \sum_n \delta_n \psi_n, \quad [84]$$

where $\sum_n \delta_n \psi_n$ a correction term to the reflection beam (also propagating in the $-x$ direction). Using the boundary conditions, we get

$$\begin{aligned} 2\psi_i - \sum_n \delta_n \psi_n &= \sum_{m=0} t_n \psi_n' \\ \sum_n \delta_n q_n \psi_n &= \sum_{n=0} t_n q_n' \psi_n' \end{aligned} \quad [85]$$

Multiplying both sides of the second equation by $\Xi_m'$ and integrating over $z$, gives

$$\sum_n \delta_n \frac{q_n}{q_m'} \int \psi_n \Xi_m' dz = t_m. \quad [86]$$

Substituting, multiplying by $\Xi_p'$ and integrating again,

$$2\left(\eta \frac{\Gamma}{t} + \gamma\left(1 - \frac{w}{2t}\theta_r\right)\right) \sum_{n=1}^{M_\Gamma} X_{n,p}' k_{p+1} - \sum_m \delta_m \int \psi_m \Xi_p' dz = \\ = \sum_m \sum_n \delta_n \left(\frac{q_n}{q_m'} \int \psi_n \Xi_m' dz\right) \int \psi_m' \Xi_p' dz \quad [87]$$

With the $\bar{\bar{X}}$ coefficient matrix being the base transform matrix defined in the appendix. This matrix links the $(\psi_n, \xi_m)$ basis, which is bi-orthogonal relative to volume integration product, with the $(\psi_n, \Xi_m)$ basis which is bi-orthogonal relative to integration over the $x$-$y$ plane. The values of $\delta_m$ are therefore determined by the following set of $M_\Gamma + 1$ linear equations,

$$\sum_n \delta_n \left(\frac{q_n}{q_p'} + 1\right) \int \psi_n \Xi_p' dz = 2(\eta + \gamma d) \sum_{n=1}^{M_\Gamma} X_{n,p}' k_{n-1}'. \quad [88]$$

This is a complete set of equations, except for the values of $\bar{\bar{X}}'$ and the $\int \psi_n \Xi_p' dz$ integrals. Notably however, the $\xi_p$ basis is very similar to the $\Xi_p$ basis (especially at larger $p$'s) since $\xi_n'$ are increasingly sine-like with growing values of $n$. Accordingly the $\int \psi_m' \xi_n' dz$ integral, for $m \neq n$, is rapidly decaying when the value of $m$ or $n$ grows and the $\bar{\bar{X}}'$ matrix can readily be approximated by an identity matrix. This is substantiated by calculating $\bar{\bar{X}}'$ explicitly

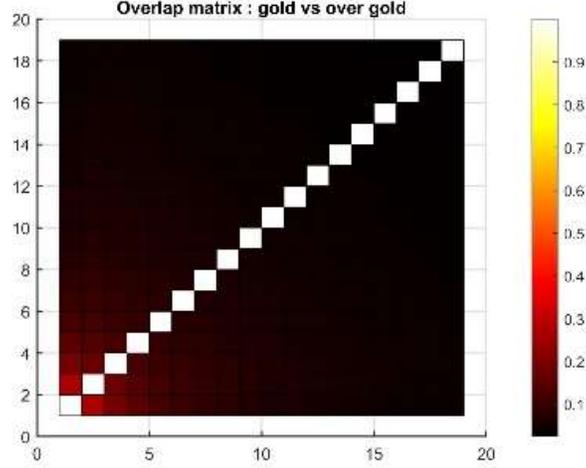

**Figure 4** - $X'_{m,n}$ matrix of the first 20 modes

This overlap matrix is clearly dominated by the delta-like response and we can therefore approximate $\bar{\bar{X}}'$ as the identity matrix and get

$$\sum_n \delta_n \left(\frac{q_n}{q'_p} + 1\right) \int \psi_n \Xi'_p dz = 2(\eta + \gamma d) k'_p. \tag{89}$$

Since the overlap integral $\eta + \gamma d$ has a minimum for $d = 0$, so will the reflection. Specifically, for the case of the specific ray-like excitation considered above, at $d = 0$ all of the terms on the R.S. scale with $\left(\frac{\Gamma}{t}\right)^{3.5}$. Therefore all of the $\delta_n$ components must also scale with $\left(\frac{\Gamma}{t}\right)^{3.5}$. Since the reflected and transmitted components is also expected to have a ray-like distribution, they should scale with $\left(\frac{\Gamma}{t}\right)^{0.5}$, but the stronger scaling we derive above implies that the total refelction indeed approaches unity when $\Gamma \to 0$.